# Installing, Running and Maintaining Large Linux Clusters at CERN

Vladimir Bahyl, Benjamin Chardi, Jan van Eldik, Ulrich Fuchs, Thorsten Kleinwort, Martin Murth, Tim Smith
*CERN, European Laboratory for Particle Physics, Geneva, Switzerland*

Having built up Linux clusters to more than 1000 nodes over the past five years, we already have practical experience confronting some of the LHC scale computing challenges: scalability, automation, hardware diversity, security, and rolling OS upgrades. This paper describes the tools and processes we have implemented, working in close collaboration with the EDG project [1], especially with the WP4 subtask, to improve the manageability of our clusters, in particular in the areas of system installation, configuration, and monitoring.

In addition to the purely technical issues, providing shared interactive and batch services which can adapt to meet the diverse and changing requirements of our users is a significant challenge. We describe the developments and tuning that we have introduced on our LSF based systems to maximise both responsiveness to users and overall system utilisation.

Finally, this paper will describe the problems we are facing in enlarging our heterogeneous Linux clusters, the progress we have made in dealing with the current issues and the steps we are taking to 'gridify' the clusters

## 1. INTRODUCTION

The LHC era is getting closer, and with it the challenge of installing, running and maintaining thousands of computers in the CERN Computer Centre.

In preparation, we have streamlined our facilities by decommissioning most of the RISC hardware, and by merging the dedicated and slightly different experiment Linux clusters into two general purpose ones (one interactive, one batch), as reported at the last CHEP[2].

Quite some progress has been made since then in the automation and management of clusters. The EU DataGrid Project (EDG), and in particular the WP4 subtask[3], has entered its third and final year and we can already benefit from the software for farm management being delivered by them. See [4] for further details. In addition, the LHC Computing Grid project (LCG)[5] has been launched at CERN to build a practical Grid to address the computing needs of the LHC experiments, and to build up the combined LHC Tier 0/Tier 1 center at CERN.

In preparing for the LHC, we are already managing more than 1000 Linux nodes of diverse hardware types, the differences arising due to the iterative acquisition cycles. In dealing with this high number of nodes, and especially when upgrading from one release version of Linux to another, we have reached the limits of our old tools for installation and maintenance. Development of these tools started more than ten years ago with an initial focus on unifying the environment presented to both users and administrators across small scale RISC workstation clusters from different vendors, each of which used a different flavour of Unix[6]. These tools have now been replaced by new tools, taken either from Linux itself, like the installation tool Kickstart from RedHat Linux or the RPM package format, or rewritten using the perspective of the EDG and LCG, to address large scale farms using just one operating system: Linux.

This paper will describe these tools in more detail and their contribution to the progress in improving the installation and manageability of our clusters. In addition, we will describe improvements in the batch sharing and scheduling we have made through configuration of our batch scheduler, LSF from Platform Computing[7].

## 2. CURRENT STATE

In May last year, the Linux support Team at CERN certified RedHat Linux 7. This certification involved the porting of experiment, commercial and administration software to the new version and verifying their correct operation. After the certification, we set up test clusters for interactive and batch computing with this new OS. This certification process took quite some considerable time, both for the users and the experiments to prepare for migration, which had to fit into their data challenges, and for us to provide a fully tailored RedHat 7.3 environment as the default in January this year. We took advantage of this extended migration period to completely rewrite our installation tools. As mentioned earlier, we have taken this opportunity to migrate, wherever possible, to the use of standard Linux tools, like the kickstart installation mechanism from RedHat and the package manager RPM, together with its package format, and to the tools that were, and still are, being developed by the EDG project, in particular by the WP4 subtask.

The EDG/WP4 tools for managing computing fabrics can be divided into four parts: Installation, Configuration, Monitoring, and Fault Tolerance. In trying to take over these ideas and tools, we first had to review our whole infrastructure with this in mind.

## 2.1. Installation

The installation procedure is divided into two main parts. The basic installation is done with the kickstart mechanism from RedHat. This mechanism allows specification of the main parameters like the partition table and the set of RPMs, and it allows the execution of arbitrary shell scripts. We are using this mechanism because it allows both a very flexible installation and automation of the installation in process. In addition, the kickstart installation can be initiated in a variety of ways; by a special floppy, by booting a special kernel or by PXE netboot, our preferred method.

In the post installation section of the kickstart mechanism, we install one startup script that is run after the next reboot, and then disabled afterwards. This script makes sure that the rest of the software is installed with RPMs, and that our configuration tool is installed and started, to configure the machine according to its intended usage. One of our goals in setting up the installation for RedHat 7 was to separate the installation and the configuration issues, which led to the principle that all software to be installed on a machine must come via an RPM package, and must then be configured by the configuration management system. Using RPM as the package manager and using its features like version control allows an easy way to update software. We have adopted this not only for the system software, which comes with the RedHat Linux distribution, but also for the software that is provided by us, and we enforce it for third parties, who want to provide software for our machines, e.g. the CASTOR software. The configuration management system, which is described in more detail in the next section, is used for configuring the software that is distributed by RPM. This has to be done for the system software to adapt it to our site-specific configuration, as well as for our own software, which allows us to provide our software in a more general way as an RPM package, to be used for other clusters as well.

The RedHat package manager RPM still lacks a good and flexible way of keeping the RPM distribution up to date on a big number of hosts, with different package lists. As an interim solution we have been using a locally-written tool called rpmupdate, which provides a very basic way of keeping the RPM list on the farm up to date. This tool has now been replaced by a new one, developed by EDG/WP4, called SPMA[8], which allows a very flexible control of the RPM list, keeping full control of all packages, and the deletion of packages if they are not explicitly configured. This is the only way to make sure that the RPM list is not outdated or otherwise modified. We have had very good experience with the first tests of the SPMA, and we are going to deploy the mechanism on all our machines in the near future.

## 2.2. Configuration

While revising our old installation of RedHat Linux 6.1, we came across more than 20 different places where configuration information for host installation was stored, ranging from databases to flat files, or hard coded in installation scripts. For configuring the hosts we still use our home made tool SUE[9], because we were not happy with the first solution adopted by WP4, LCFG[10]. But we have decided to use the configuration database developed by the configuration subtask of WP4, PAN[11], which is a very flexible and sophisticated tool for describing host configuration. The host configuration is described in a language, called PAN, and it is compiled into an XML file for each host, and this information is both made available through an API and cached on the target machine. It is one of our major tasks to migrate the configuration information from all the different historical places into this unique one. We have already made this information available on the node itself through a common interface called CCConfig.pm, which is a PERL module, because almost the whole configuration code is written in PERL language. This interface can be seen as the high level API of the configuration information.

Each host is described in PAN according to a global schema, which is a tree-like structure of host information. Note, however, that the exact details of the global schema and its description in the High Level Description Language of PAN are still evolving as they get more heavily used.

One main branch of this tree-like structure comprises the software components, or features, as they are called in SUE, which will have to be rewritten as we go from SUE to the new tool currently being developed in WP4, called the 'Node Configuration Manager, NCM' [12].

## 2.3. Monitoring

In parallel, but independent from the development of the installation and configuration of our farms, the monitoring was completely rewritten. Here the difficulty was that we had to replace the old tools not only on the new platform RedHat Linux 7, but simultaneously also on the old RedHat 6.1 nodes, as well as on other platforms, like SUN running Solaris. In addition, the monitoring had to run on machines other than standard compute servers, such as disk or tape servers, which were still being managed and installed in a different way. Therefore the requirements were much broader for the monitoring than for installation of configuration. Again, though, we could benefit from EDG/WP4 developments, in this case from the monitoring subtask. Now the whole monitoring system on the clients is replaced by WP4 monitoring. Investigations on this part of the monitoring have not been concluded yet. Similarly, we are currently still investigating alternative solutions for collecting the monitoring information from each node and storing this in a relational database. Currently, we are storing the monitoring information in an ORACLE database.

## 2.4. Fault Tolerance

The fault tolerance subtask of WP4 is responsible for taking actions on detected system errors. This is a very sophisticated goal and the investigations are still ongoing. Consequently, we have not yet decided if we will implement this solution. For the moment there is no attempt at automated corrective action. Instead, operators follow standard procedures in response to alarms from the monitoring system. This may change in the future, as the details of an effective automated system are not at all clear to us at present.

## 2.5. Collaboration with the EDG project

As already pointed out in several previous paragraphs, we have a close collaboration with the EDG project, and in particular with the WP4 subtask. This partly reflects the fact that some members of the operation team are active contributors to WP4, matching the EU funded effort. There is a complementary effect: We can directly influence the work of this subtask by giving input for further developments on one hand, and on the other hand, the WP4 subtask has an excellent 'testbed', in that they can not only test, but also use their application on such a big farm as the one in the CERN computer centre. This is a very fruitful collaboration, and we hope the work can be continued after the EDG project finishes at the end of this year.

## 3. MAINTENANCE OF THE CLUSTERS

Beside the new ways of installing, configuring and monitoring the clusters, as described in the previous sections, we have also reviewed our way of maintaining them. Once the number of machines you have to maintain increases beyond 1000 or so, you cannot rely any more on centralised tools that require client machines to be up and running at a given time to allow configuration changes. There are always machines that are broken, or in a state in which they cannot reconfigure due to some problem or other. To avoid inconsistencies we have designed our tools such that machines can be reinstalled at any time, with a reinstall returning the machine to the same state as it was in before the reinstall or failure. This is extremely constraining, but extremely important, since machines can and do fail at any time. One important issue here is that the installation procedure has to be completely automated, otherwise the effort to reinstall a machine is too high, and needs expert intervention. In addition, this approach needs to have all the configuration information necessary to set up a machine to be stored outside the node itself, otherwise it would be lost during a reinstallation. The converse must be true also. If a live machine is 'updated', e.g. by changing the configuration or the RPM packages, it should end up with the same setup as a machine that has been reinstalled, and hence machine changes should result only from changes in the central configuration database. In addition, machines that are down for some time, or simply in a bad (software) state, should be reinstalled to catch up with the latest setup, and the configuration tool has to be idempotent to allow multiple runs of it without disturbing the system.

## 4. OTHER IMPROVEMENTS

In addition to the above-mentioned improvements using EDG tools, we have made some other developments, which were needed in order to handle large numbers of computers. All these developments were to automate the installation and the maintenance. Most of these tools are already in use but are still being improved, because we learn a lot as we extend to even bigger numbers of machines. They are described in more detail in the next subsections.

## 4.1. Secure installations

One major problem of any automated installation is the question of security: How do I make sure that secure information that is needed during the installation, like SSH host keys or the root password, is put onto the machine in a secure way? This includes the storage of this information outside the node before the installation, as well as the transport onto the node. We have solved one part of the problem by creating a GPG key[13] for every host in our clusters. This GPG key is used to encrypt valuable information such as that mentioned above. By doing so, we can put this encrypted information onto our installation server without any further security measures, because the information can only be used by someone who has the private part of the GPG key, which is only the server that generated it and the client itself. This leads to the obvious question: How is the private part of the GPG key delivered to the machine to be installed? One way is to put it onto the floppy disk with which the host is installed. This way the installation can be done even on a non-trusted network. We proceed a different way because we do not install our machines using a floppy, but with net-boot. This way the GPG key is transported to the client in a very early stage of the installation via SCP, whereby the server has to trust the network connection. There is nothing on the client that can be used for authentication but the IP-address. Security is enhanced by allowing this secure copy only during a very short time window, which is be opened on the server just prior to install. Recreating a GPG key pair for each host on each reinstallation increases the security further.

## 4.2. Intervention Rundown

One big managerial problem arises if it is necessary to shutdown or reboot a whole cluster of machines, especially batch nodes. On batch nodes, you have to stop the batch system, wait until the last job has finished, and only then can you do the intervention. Depending on the maximal runtime of the batch jobs, this can take, in our case, up to one week until the last job has finished, whilst other nodes are empty after only a few minutes. This can lead to a lot of lost compute time on your cluster if you wait for the last node! To avoid this, we are currently testing a system that runs on individual hosts, disables the batch system on this host in a way that no new job is scheduled, and, as soon as the host is drained, the intervention rundown starts. This can be something like a reboot to install a new kernel, or a reinstall if changes have to be made that are easiest done by a reinstall. The Intervention Rundown takes care of all the necessary steps, e.g. sending emails and disabling the monitoring system. The system works on interactive systems as well in which case we disable the system for new logins, give the already logged in users a configurable grace time to finish their work and do the intervention rundown afterwards.

### 4.3. Server Cluster

We have concentrated the service functions to run our clusters on a special cluster, called Server Cluster. This cluster provides all necessary functionality to the clients, including the software repository for RPM, the GPG key server and the configuration information through XML files. When this server cluster is fully functioning, we will serve everything on this cluster through a web server. We see several advantages of this: The HTTP protocol is one of the most used these days, which means well tested and very scalable solution exist already. We run a web server on each node of the server cluster, and we access these servers through a kind of 'round-robin' of a DNS name switch, which allows this web services to be highly available—again, a common and well tested setup for http servers. No special hardware is required and the cluster is easily scalable, through the addition of 'off the shelf' Linux boxes, to serve for O(10,000) nodes in the future.

### 4.4. Notification Mechanism

Another new service running on our server cluster is the notification mechanism. This mechanism allows clients to subscribe to a server for a special 'tag', and the server will notify all subscribed clients, when somebody wants to notify for this special tag. A tag is usually a simple word like 'rpmupdate', or 'confupdate'. This procedure is now used for RPM and configuration updates. We do not run such tasks regularly by a daily CRON job anymore, but only on notification. This allows us to have full control of the updates.

### 5. THE BATCH SCHEDULER

We are using LSF from Platform Computing[6] for our batch scheduler. The installed version is 4.2, and we are investigating version 5.1 at the moment. Within the last year, we have made two major changes. First, we have stopped using the multi cluster option. This option allows different clusters to be run independently, with communications routed only through their master hosts. This is a very nice feature if you have a lot of cross-linked clusters, as we had in the past. However, information was passed only partially from one cluster to the other. For example, the reason why a job was pending was not at all obvious to the user. As we have reduced the number of clusters to effectively two, an interactive and a batch one, there was no particular need for using this option and much to be gained in terms of overall clarity for the users by dropping it. The second change was done by the introduction of the fairshare mechanism of LSF. This allows on one hand to guarantee a fixed percentage (share) of the whole batch capacity for each experiment when it is needed, but on the other hand, others can use the capacity when it is not needed. This has led to a much better utilization of the farms, avoiding pending jobs so long as there is free capacity and no other limits have been reached. This second change was well received by the experiments because it has increased their usable capacities.

### 6. LCG

The LHC Computing Grid, LCG, is the project that was started to deal with the computing requirements for the LHC. CERN, as the host site of the LHC, will run the Tier 0 center, and in addition, we will have a Tier 1 center to cater for the needs of physicists based at CERN and those in countries without a dedicated Tier1 centre. See e.g. [14] for details. Our current clusters of LXPLUS and LXBATCH will evolve into the computing capacity of the Tier 1 of CERN in the future. As an initial step, the first prototype of the LCG software, LCG0, has been released, and deployed on some test nodes, to see the impact of this new software on our current setup, and to solve problems encountered. It is planned to setup a large fraction, if not all, of our current farms with the LCG1 release that is expected this summer. Unfortunately this initial software places some requirements on our setup, such as use of NFS as a network file system and direct WAN access for each compute node that are incompatible with our plans and constraints for the long term. Solving these problems will be one of the big tasks for us and for the LCG team this year.

### 7. FUTURE PLANS

As described in this paper, we have made big progress in our installation and maintenance procedures for Linux in the last year when we went from RedHat 6.1 to RedHat 7.3. We will switch off the old OS version by this summer. We will continue to work on the improvements on our procedures. Beside the continuation on the projects described above, one main issue for the future is to replace our current configuration tool by the tool from WP4, the NCM. In addition, the range of our installation and configuration tools has now been extended to other types of cluster, e.g. disk and tape servers, as well. This necessitates broadening the operational area of these tools to meet the special requirements of these different clusters. As an example a disk server machine has lots of disks attached whose configurations have to be stored, and which should not be deleted during a system installation, whereas on compute servers we only have one or two system disks, which are normally simply reformatted. Other examples are special service clusters, e.g. a central CVS server cluster or a special build cluster, for regular experimental software compilations. These clusters have to be treated differently in terms of user access, RPM package list, etc.

Having new tools at hand it is now easy to integrate these new clusters into our installation and maintenance procedures.